\documentclass[aps,pre,twocolumn,showpacs,letterpaper, preprintnumbers]{revtex4}
\usepackage{amsmath, amsthm, amssymb}
\usepackage{graphicx}

\DeclareMathOperator{\tr}{tr}

\newcommand{\sop}[1]{\ensuremath{\mathcal{#1}}} 

\newcommand{\Env}{\ensuremath{^{\text B}}}
\newcommand{\Sys}{\ensuremath{^{\text S}}}

\newcommand{\env}{\ensuremath{_{\text B}}}
\newcommand{\sys}{\ensuremath{_{\text S}}}

\newcommand{\kB}{k_\mathrm{B}}

\begin{document}

\title{On the Quantum Jarzynski Identity}
\author{Gavin E. Crooks}
\email{gecrooks@lbl.gov}
\affiliation{Physical Biosciences Division, Lawrence Berkeley National Laboratory, Berkeley, California 94720}

\date{\today}

\begin{abstract}
In this note, we will discuss how to compactly express and prove the Jarzynski identity for an open quantum system with dissipative dynamics.   We will avoid explicitly measuring the work directly, which is tantamount to continuously monitoring the system, and instead measure the heat flow from the environment. We represent the measurement of heat flow with Hermitian map superoperators  that act on the system density matrix.  Hermitian maps provide a convenient and compact representation of sequential measurement and correlation functions. 
\end{abstract}
	
\preprint{LBNL-62801}
\maketitle


\section*{Introduction}
When a classical system in thermal equilibrium is driven from that equilibrium by an external perturbation, then the irreversible work of that process is related to the system's free energy change  by Jarzynski's equality~\cite{Jarzynski1997a, Crooks1998,Hummer2001a,Liphardt2002}.
\begin{equation}
\left \langle e^{-\beta W} \right\rangle = \int p(W) \,e^{-\beta W} dW
= e^{-\beta \Delta F}.
\label{jarzynski}
\end{equation}
Here,  $p(W)$ is the probability distribution of  work $W$ done on the system, $\beta=1/\kB T$ is the inverse temperature $T$ of the environment in natural units, ($\kB$ is Boltzmann's constant) and $\Delta F$ is the Helmholtz free energy change of the system. 
In other words, a Boltzmann weighted average of the irreversible work recovers the equilibrium free energy difference from a non-equilibrium transformation.

\nocite{Tasaki2000,Kurchan2001}
The generalization of the Jarzynski identity to quantum dynamics~\cite{Tasaki2000,Kurchan2001, Mukamel2003a,Chernyak2004, DeRoeck2004, Jarzynski2004a, Monnai2005, DeRoeck2006, Esposito2006,Chernyak2006b, Esposito2007, Talkner2007a, Talkner2007b}, although technically straightforward, is not without some interesting conceptual complications.  The total change in energy of the system $\Delta E= Q + W$ is equal to the work $W$ applied via the time dependent perturbation plus the flow of heat $Q$ from the environment. 
In a classical system, we can continuously measure the energy of the system and the ebb and flow of heat and work. However, continuous measurement is not possible in quantum mechanics without severally disturbing the dynamics of the system. We should not attempt to continuously monitor the system~\cite{Yukawa2000}, directly measure the work~\cite{Engel2007} or radically redefine work and heat~\cite{Yukawa2000, Allahverdyan2005, Engel2007}.

The simplest solution to this apparent dilemma is to disallow heat flow during the experiment~\cite{Tasaki2000, Kurchan2001}. The system is initially in thermal equilibrium with the environment, but is decoupled and isolated from the environment during the perturbation of the system. The work is then the difference in energy of the system between the beginning and end of the experiment. 

A more satisfactory solution is to realize that although we cannot continuously measure the work  or system energy, we can measure the heat flow from the environment~\cite{DeRoeck2004}. The total system is divided into a system-of-interest and a thermal bath. This environment is large, rapidly decoheres and always remains at thermal equilibrium, uncorrelated and unentangled with the system.  Consequently, we can measure the change in energy of the bath (i.e.~$-Q$) without further disturbing the dynamics of the system. Essentially, we reexpress the open-system quantum Jarzynski identity as
\begin{equation}
\left \langle e^{-\beta W} \right\rangle = \left \langle e^{-\beta E_f} e^{+\beta Q} e^{+\beta E_i} \right\rangle =  e^{-\beta \Delta F}.
\label{quantumjarzynski}
\end{equation}

In this note I will discuss how to represent the heat flow and quantum Jarzynski identity with Hermitian maps, generalized measurement superoperators that act on the system density matrix alone.


\section*{Background: Quantum Dynamics of Open Systems}

The mixed state of a quantum system that interacts  with the 
environment can be described by a density matrix, $\rho$, a positive operator 
with unit trace. For an equilibrated system with Hamiltonian $H$ interacting weakly with a thermal  bath of temperature $T$ the equilibrium density matrix is
\begin{equation}
	\rho_{\text{eq}} = \frac{e^{-\beta H} }{\tr e^{-\beta H} } =  \frac{e^{-\beta H} }{Z} =  e^{+\beta F -\beta H} 
\label{densitymatrix}
\end{equation}
where $\beta = 1/k_{\text B} T$, $k_{\text B}$ is Boltzman's constant, $Z= \tr \exp(-\beta H)$ is the partition function, and $F=-\frac{1}{\beta} \ln Z$ is the Helmholtz free energy of the system.

The dynamics of an open quantum system  can be described by a quantum operation $\rho'=\sop{S}\rho$,  a linear, trace preserving, complete positive map of operators~\cite{Kraus1983,Schumacher1996,Caves1999,Nielsen2000}. 
Any such complete, positive superoperator  has operator-sum (or Kraus) representations,
\begin{equation}
	      \sop{S} \rho \equiv \sum_{\alpha} A_{\alpha} \rho A^{\dagger}_{\alpha}
\label{kraus}	      
\end{equation}
Conversely, any operator-sum represents a  complete, 
positive superoperator. The collection $\{A_{\alpha}\}$ are known as Kraus operators. 
The superoperator is trace preserving, and therefore conserves probability, if $\sum_{\alpha} A^{\dagger}_{\alpha} A_{\alpha} = I$, where $I$ is the identity operator. In the simplest case, the dynamics of a isolated quantum system is described by a single unitary operator $U^{\dagger} = U^{-1}$. 

We are interested in the dynamics of a quantum system with a time dependent Hamiltonian, weakly coupled to an extended, thermal environment.  Let the total Hamiltonian of the combined system be
\begin{equation}
H= H\Sys(t) \otimes  I\Env+ I\Sys\otimes H\Env + \epsilon H^{\text {int}} ,
\label{hamiltonian}
\end{equation}
 where $I\Sys$ 
and $I\Env$ are system and bath identity operators, $H\Sys$ is the time dependent Hamiltonian of the system,  $H\Env$ is the bath Hamiltonian, $H^{\text {int}}$ is the bath-system interaction Hamiltonian and  $\epsilon$ is a small coupling constant.

We assume that initially the system and environment are uncorrelated and that the initial combined state is  $\rho\Sys \otimes \rho\Env_{\text{eq}}$, where  $\rho\Env_{\text{eq}}$  is the thermal equilibrium density matrix of the bath.
We can derive a quantum operator description of the system dynamics by following the unitary dynamics of the combined total system for a finite time and then measuring the final state of the environment.
\begin{align}
\sop{S}(s,t) \rho\Sys  &= \tr\env U
 [\rho\Sys \otimes \rho\Env] U^{\dagger} 
\label{unitary}
 \\
&= 
\sum_{f} 
  \langle b_{f} | 
  U
  \left(
  \rho\Sys  \otimes
	\left[\sum_{i}\frac{e^{-\beta E\Env_{i}}}{Z\Env}
	| b_{i}\rangle\langle b_{i} | \right] 	
  \right) 
  U^{\dagger}
| b_{f} \rangle
\nonumber \\
&= \sum_{if} 
\frac{e^{-\beta E\Env_{i}}}{Z\Env}
\langle b_{f} |   U |  b_{i} \rangle	
\;  \rho\Sys \;
\langle b_{i} | U^{\dagger} | b_{f} \rangle	
 \nonumber
\end{align}
Here,
$U$ is the unitary evolution operator of the total system 
\begin{equation}
U= \exp\left( -\frac{i}{\hbar} \int_s^t H(\tau) \,d\tau \right) \,,
\end{equation}
 $\tr\env$ is a partial trace over the bath degrees of freedom, 
$\{E\Env_{i}\}$ are the energy eigenvalues and $\{| b_{i}\rangle\}$ are the orthonormal energy eigenvectors of the bath, and $Z\Env$ is the bath partition function. For simplicity, and without loss of generality, we assume that the bath energy states are non-degenerate.

 It follows from the last line of Eq.~(\ref{unitary}) that the Kraus operators for this dynamics are
\begin{equation}
A_{ij} = \frac{e^{-\frac{1}{2}\beta E\Env_{i}}} {\sqrt{Z\Env }} \langle b_{j} | 
U | b_{i} \rangle
\label{thermostattedKraus}
\end{equation}

We further assume that the environment is large, with a characteristic relaxation time short compared with the relevant bath-system interactions and that the system-bath coupling $\epsilon$ is small. Thus the environment remains very near thermal equilibrium,  unentangled and uncorrelated with the system. Consequently, the system dynamics of each consecutive time interval can be described by superoperators derived as in Eq.~(\ref{unitary}), which can then be chained together to generate a quantum Markov chain. 
\begin{equation}
\rho(t) = \sop{S}(t-1,t) \cdots   \sop{S}(s+1,s+2)\,  \sop{S}(s,s+1)\, \rho(s)
\end{equation}
In the limit of small time interval we obtain a continuous time quantum Markovian dynamics, 
\begin{equation}
\rho(t) = \exp\left(\int_s^t \sop{L}(\tau) d\tau \right) \rho(s)
\end{equation}
where $\sop{L}$ is the Lindbladian 
superoperator~\cite{Nielsen2000}.


\section*{Hermitian Maps and Sequential Measurements}

A measurement of a quantum system can be characterized by a collection of measurement operators $\{A_{\alpha}\}$, where $\sum_{\alpha} A^{\dagger}_{\alpha} A_{\alpha}=I$, and associated real valued measurement results, $a_{\alpha}$. For example, the Hermitian operator $H=H^{\dagger}$ of a standard von Neumann measurement can be decomposed into a collection of eigenvalues $h$ and orthonormal projection operators $P_h$,  such that $H = \sum_h hP_h$. More generally, the measurement operators of a POVM (Positive Operator Valued Measure) need not be projectors nor orthonormal~\cite{Nielsen2000}.

The probability of observing the $\alpha$th outcome is
\begin{equation}
p_{\alpha}  = \tr A_{\alpha}  \rho A^{\dagger}_{\alpha} 
\end{equation}
and the state of the system after this particular interaction is
\begin{equation}
\rho'_{\alpha} = 
\frac{A_{\alpha}  \rho A^{\dagger}_{\alpha} }{\tr A_{\alpha}  \rho A^{\dagger}_{\alpha} }
\,.
\end{equation}
The overall effect of the dynamics, averaging over different interactions, is the full quantum operation, Eq.~(\ref{kraus}). 

Rather than simple representing the \emph{effect} of the measurement with the appropriate quantum operation, we can represent the  \emph{effect} and  \emph{result} of the measurement using a Hermitian map superoperator~$\sop{A}$ :
\begin{equation}
   \sop{A} \rho = \sum_{\alpha} a_{\alpha} A_{\alpha} \rho 
   A^{\dagger}_{\alpha}
\end{equation}
Note that this operator-value-sum cannot, in general, be recast as an operator-sum, since the measurement values  $\{a_{\alpha}\}$ may be negative. An operator-value-sum maps Hermitian operators to Hermitian
operators  ($H=H^{\dagger}$) ,
\begin{equation}
   [\sop{A} H]^{\dagger} = [\sum_{\alpha} a_{\alpha} A_{\alpha} H 
   A^{\dagger}_{\alpha}]^{\dagger} 
	 = \sop{A} H^{\dagger} = \sop{A} H
\end{equation}
 Conversely, any Hermitian map has an operator-value-sum
representation~\cite{Rungta2001}.

Hermitian maps provide a particularly  compact and convenient  representation 
of  sequential measurements and correlation functions.
Let the Hermitian map $\sop{A}$ representing a measurement at
time 0, $\sop{B}$ a different measurement of the same system at  time $t$, and the quantum operation $S_t$ represent the system evolution
between these two measurements. 
 The expectation value of a single measurement  is 
\begin{equation}
\langle a \rangle = \tr \sop{A} \rho 
= \sum_{\alpha} a_{\alpha} \tr A_{\alpha} \rho A^{\dagger}_{\alpha} 
= \sum_{\alpha} p({\alpha}) a_{\alpha}
\end{equation}
and the correlation function $\langle b(t) a(0)\rangle$ can be expressed as
\begin{align}
\langle b(t) a(0)\rangle 
&= \tr \sop{B} \sop{S}_t \sop{A} \rho(0) 
\nonumber \\
&= \sum_{\alpha \beta}  a_{\alpha} b_{\beta} 
\tr B_{\beta}  \left[ S_t[ A_\alpha \rho(0) A^{\dagger}_\alpha ] \right] B^{\dagger}_\beta 
\nonumber \\
&= \sum_{\alpha \beta}  p(\alpha,\beta)  a_{\alpha} b_{\beta}
\,.
\end{align}
(Note that expressions such as $\tr AB\rho$, where $A$ and $B$ are 
Hermitian operators, often appear in perturbation expansions and are frequently referred to as quantum correlation functions. However, since $AB$ is not in general Hermitian these expressions do not directly represent a physical measurement.)

Here we establish that, just as every Hermitian operator represents some measurement on the Hilbert space of pure states, every Hermitian map can be associated with some measurement on the Liouville space of mixed states. Suppose that we have already decomposed the Hermitian map $\sop{A}$ into a value-operator-sum with values $\{a_{\alpha}\}$ and operators $\{A_{\alpha}\}$. Probability conservation requires that  $\sum_{\alpha} A^{\dagger}_{\alpha} A_{\alpha} = I$. If this condition is not met we can supplement the Kraus operators with an additional operator whose corresponding measurement value is zero. Note that $A^{\dagger}_{\alpha} A_{\alpha}$ is a positive operator and consequently $B^{\dagger}B = I - \frac{1}{c} \sum_{\alpha} A^{\dagger}_{\alpha} A_{\alpha}$ is also a positive operator provided that $c$ is a  real number larger than the largest eigenvalue of  $\sum_{\alpha} A^{\dagger}_{\alpha} A_{\alpha}$. Therefore, we can rescale the measurement outcomes $\{ c a_{\alpha} \}$  and Kraus operators $\{ \frac{A_{\alpha}}{\sqrt{c}}\}$, append the additional operator $B$ with measurement outcome $0$, and associate the superoperator $\sop{A}$ with the measurement 
\begin{equation}
\sop{A} \rho = 0 \, B \rho B^{\dagger} +
\sum_i  c a_{\alpha} \frac{A_{\alpha}}{ \sqrt{c} } \,\rho\, \frac{A_{\alpha}^{\dagger}}{ \sqrt{c} } 
\end{equation}
Note that the decomposition of a Hermitian map into an operator-value-sum representation is not unique~\cite{Nielsen2000,Rungta2001}.

\section*{Measurements of Heat Flow}

We can now construct a Hermitian map representation of heat flow, under the  assumptions  that the bath and system Hamiltonians are constant during the measurement procedure and that the bath-system coupling is small. We construct a measurement on the total system
and then project out the bath degrees of freedom, leaving a Hermitian map superoperator that acts on the system density matrix alone. 

The full, explicit measurement is
\begin{align*}
\left\langle e^{+\beta Q}  \right\rangle = 
\overbrace{\sum_{if}}^{\text{h}} 
 \overbrace{
  e^{ -\beta (E\Env_f -E\Env_i)}
 }^{\text{g}} 
\,
\overbrace{ \tr\sys \phantom{^{\dagger}}}^{\text{f}} 
\overbrace{ \tr\env \phantom{^{\dagger}}}^{\text{e}} 
\,
\overbrace{ [ I\Sys \otimes | b_{f}\rangle\langle b_{f} |] }^{\text{d}} 
\,
\\
\overbrace{U\phantom{^{\dagger}}}^{\text{c}}\,
\overbrace{
[ I\Sys \otimes | b_{i}\rangle\langle b_{i} | ]}^{\text{b}}
\,
\overbrace{
[  \rho\Sys \otimes \rho_{\text{eq}}\Env]
}^{\text{a}}
\,
\overbrace{[ I\Sys \otimes | b_{i}\rangle\langle b_{i} |]}^{\text{b}}
\,
\overbrace{U^{\dagger}}^{\text{c}}
\,
\\
\overbrace{[ I\Sys \otimes | b_{f}\rangle\langle b_{f} | ]}^{\text{d}}
\nonumber
\end{align*}
We start with a composite system consisting of the bath, initially in thermal equilibrium, weakly coupled to the system (a). We measure the initial energy eigenstate of the bath (b), allow the system and bath to evolve together for some time (c), and then measure the final energy eigenstate of the bath (d). The trace over the bath degrees of freedom (e) yields the final, unnormalized system density matrix, whose trace in turn (f)  gives the probability of observing the given initial and final bath energy eigenstates. We then multiple by the Boltzmann weighted heat (g) and sum over all initial and final bath states (h) to obtain the desired average Boltzmann weighted heat flow.

The sum over initial states can be split into separate sums on the left and right projectors, since the bath is initially diagonal. Similarly, the sum over the final states can be split into separate sums on the right and left due to the final trace over bath degrees of freedom. As a consequence, we can rewrite the previous expression using the bath Hamiltonian. 
\begin{align*}
=\tr\sys \tr\env 
[ I\Sys \otimes e^{-\frac{\beta}{2}H\Env}] 
\,U\,
&[ I\Sys \otimes e^{+\frac{\beta}{2}H\Env}] 
\,  [ \rho\Env \otimes \rho_{\text{eq}}\Env]\,\cdot
\nonumber \\ & 
 \cdot [ I\Sys \otimes e^{+\frac{\beta}{2}H\Env}] 
\,U^{\dagger}\,
[ I\Sys \otimes e^{-\frac{\beta}{2}H\Env}] 
\nonumber
\end{align*}
We replace the bath Hamiltonian by $I\Sys\otimes H\Env = H -  H\Sys(t) \otimes  I\Env - \epsilon H^{\text {int}} $ [Eq.~(\ref{hamiltonian})]. The total Hamiltonian commutes with the unitary dynamics and cancels, and we can discard the interaction Hamiltonian in the small coupling limit.
\begin{align}
&\approx
\tr\sys \tr\env 
[ e^{+\frac{\beta}{2}H\Sys}  \otimes I\Env ] 
\,U\,
[ e^{-\frac{\beta}{2}H\Sys}  \otimes I\Env ] 
\,[ \rho\Sys \otimes \rho_{\text{eq}}\Env]\,\cdot
\nonumber \\ & \nonumber \qquad\qquad\qquad\qquad\qquad\quad \cdot
 [ e^{-\frac{\beta}{2}H\Sys}  \otimes I\Env ] 
\,U^{\dagger}\,
[ e^{+\frac{\beta}{2}H\Sys}  \otimes I\Env  ] 
\end{align}
We now collect terms acting on the bath or system alone
\begin{equation*}
 =
\tr\sys \, 
e^{+\frac{\beta}{2}H\Sys}
 \Big[\tr\env 
\,U\,
\big[ [e^{-\frac{\beta}{2}H\Sys}  \, \rho\Sys\, e^{-\frac{\beta}{2}H\Sys}] \otimes  
\rho_{\text{eq}}\Env \big]
\,
U^{\dagger} 
\,
\Big] e^{+\frac{\beta}{2}H\Sys}
\end{equation*}
and recover the Kraus operators $\{A_{\alpha}\}$ describing the reduce dynamics of the system, Eq.~(\ref{thermostattedKraus}).
\begin{equation*}
=\tr\sys \,  \sum_{\alpha} 
e^{+\frac{\beta}{2}H\Sys} A_{\alpha} e^{-\frac{\beta}{2}H\Sys}  
\,  \rho\Sys\, 
e^{+\frac{\beta}{2}H\Sys}  A_{\alpha}^{\dagger} e^{-\frac{\beta}{2}H\Sys}
%
%
\end{equation*}
Finally, we have found that the average Boltzmann weighted heat flow can be represented by 
\begin{equation}
 \left\langle e^{+\beta Q}  \right\rangle = \tr \sop{R}^{-1} \sop{S} \sop{R} \,\rho\Sys
\end{equation}
where $\sop{S}$ is the reduced dynamics of the system and the Hermitian map measurement superoperator $\sop{R}$ is 
\begin{equation}
\sop{R}_t \rho= 
e^{-\frac{\beta}{2} H_t} 
\; \rho \;
e^{-\frac{\beta}{2} H_t} 
\label{R}
\end{equation}
The paired Hermitian map superoperators act at the beginning and end of a time interval and measure the change in system energy during that interval. This does not disturb the system beyond the disturbance already induced by coupling the system to the environment.

\section*{Quantum Jarzynski Identity}
We are now in a position to derive the quantum Jarzynski identity [Eq.~(\ref{quantumjarzynski})] using Hermitian maps and the quantum operator formalism. We split the total experimental time into a series of discrete intervals, labeled by the integer $t$. The system Hamiltonian is fixed within each time interval, and only changes in discrete jumps at the interval boundaries~\cite{Crooks1998}. We can therefore measure the heat flow by wrapping the superoperator time evolution of each time interval $\sop{S}_t$ with the corresponding Hermitian map measurements $\sop{R}_t^{-1} \sop{S}_t \sop{R}_t$. Similarly, we can represent the measurement of the Boltzmann weighted change in energy of the system  with $\langle\exp(-\beta\Delta E\rangle = \tr \sop{R}_\tau \sop{S} \sop{R}^{-1}_0$. 

The average Boltzmann weighted work of a driven, dissipative quantum system can therefore be compactly expressed as
\begin{equation}
\left\langle e^{-\beta W}  \right\rangle =
 \tr \big[
\sop{R}_{\tau}
\left( \prod_t 
[\sop{R}^{-1}_{t} \sop{S}_t \sop{R}_{t} ]
\right)
 \sop{R}^{-1}_{0}
\rho^{\text{eq}}_0
\big]
\end{equation}
where $\rho^{\text{eq}}_t$ is the system equilibrium density matrix with system Hamiltonian $H_t\Sys$.

Due to the structure of the energy change Hermitian maps $\sop{R}$ [Eq.~(\ref{R})] and the equilibrium density matrix [Eq.~(\ref{densitymatrix})] this product of superoperators telescopes, leaving only the free energy difference between the initial and final equilibrium ensembles. 
\begin{align}
\left\langle e^{-\beta W}  \right\rangle &\\ 
 \nonumber=  \tr & \big[ 
\sop{R}_{\tau}
[\sop{R}^{-1}_{\tau} \sop{S}_\tau \sop{R}_{\tau} ]
 \cdots 
[\sop{R}^{-1}_{2} \sop{S}_2 \sop{R}_{2} 
][
\sop{R}^{-1}_{1} \sop{S}_1 \sop{R}_{1} 
]
 \sop{R}^{-1}_{0}
\rho^{\text{eq}}_0 \big]
\nonumber
\\
 =  \tr & \big[ 
\sop{R}_{\tau}
[\sop{R}^{-1}_{\tau} \sop{S}_\tau \sop{R}_{\tau} ]
 \cdots 
[\sop{R}^{-1}_{2} \sop{S}_2 \sop{R}_{2} 
][
\sop{R}^{-1}_{1} \sop{S}_1 \sop{R}_{1} 
]
\frac{I}{Z(0)}
\big]
\nonumber
\\
 =  \tr & \big[ 
\sop{R}_{\tau}
[\sop{R}^{-1}_{\tau} \sop{S}_\tau \sop{R}_{\tau} ]
 \cdots 
[\sop{R}^{-1}_{2} \sop{S}_2 \sop{R}_{2} 
]
\sop{R}^{-1}_{1} \sop{S}_1 \rho^{\text{eq}}_1
\frac{Z(1)}{Z(0)}
\big]
\nonumber
\\
 =  \tr & \big[ 
\sop{R}_{\tau}
[\sop{R}^{-1}_{\tau} \sop{S}_\tau \sop{R}_{\tau} ]
 \cdots 
[\sop{R}^{-1}_{2} \sop{S}_2 \sop{R}_{2} 
]
\sop{R}^{-1}_{1} \rho^{\text{eq}}_1
\frac{Z(1)}{Z(0)}
\big]
\nonumber
\\ = \phantom{\tr}&
 \frac{Z(\tau)}{ Z(0)} = \exp\{ -\beta \Delta F \}
\nonumber
\end{align}

In the limit of small time intervals, we can express the quantum Jarzynski identity in the continuous time 
Llindblad form.
\begin{align}
\left\langle e^{-\beta W}  \right\rangle 
&=\tr \sop{R}(t) 
\exp\left\{
\int_0^{\tau} R(t)^{-1} \sop{L}(t) \sop{R}(t) dt\right\}
\sop{R}(0)^{-1}
\rho^{\text{eq}}_0
\nonumber \\&
= e^{-\beta\Delta F}
\end{align}


Financial support was provided by the DOE/Sloan Postdoctoral Fellowship in Computational Biology; by the Office of Science, Biological and Environmental Research, U.S. Department of Energy under Contract No. DE-AC02-05CH11231; and by California Unemployment Insurance.

\bibliography{GECLibrary}	

\end{document}